\documentclass[12pt]{article}
\usepackage{amssymb,latexsym,amsmath,graphicx}
\begin{document}
\newtheorem{theorem}{Theorem}
\newtheorem{problem}[theorem]{Problem}
\newtheorem{proposition}[theorem]{Proposition}
\newtheorem{definition}[theorem]{Definition}
\newtheorem{lemma}[theorem]{Lemma}
\newtheorem{corollary}[theorem]{Corollary}
\newtheorem{remark}[theorem]{Remark}
\newtheorem{conjecture}[theorem]{Conjecture}
\def\CC{\Bbb{C}}
\def\RR{\Bbb{R}}

\author{Gaik Ambartsoumian and Peter Kuchment\\
Mathematics Department\\
Texas A \& M University\\
College Station, TX 77843-3368\\
kuchment@math.tamu.edu, haik@tamu.edu}
\title{On the injectivity of the circular Radon transform}
\date{}
\maketitle
\begin{abstract}
The circular Radon transform integrates a function over the set of
all spheres with a given set of centers. The problem of
injectivity of this transform (as well as inversion formulas,
range descriptions, etc.) arises in many fields from approximation
theory to integral geometry, to inverse problems for PDEs, and
recently to newly developing types of tomography. A major
breakthrough in the $2D$ case was made several years ago in a work
by M.~Agranovsky and E.~T.~Quinto. Their techniques involved
microlocal analysis and known geometric properties of zeros of
harmonic polynomials in the plane. Since then there has been an
active search for alternative methods, especially the ones based
on simple PDE techniques, which would be less restrictive in more
general situations. The article provides some new results that one
can obtain by methods that essentially involve only the finite
speed of propagation and domain dependence for the wave equation.
\end{abstract}

\section{Introduction}

The circular Radon transform integrates a function over the set of
all spheres with a given set of centers. Such transforms have been
studied over the years in relation to many problems of
approximation theory, integral geometry, PDEs, sonar and radar
imaging, and other applications (\cite{A}--\cite{And},
\cite{CH}--\cite{FPR}, \cite{GGG2}--\cite{Gi},
\cite{John}--\cite{LQ}, \cite{Natt2001}--\cite{Qrange},
\cite{MXW}--\cite{XWAK}). Although significant progress has been
achieved, some related analytic problems have proven to be rather
hard and remain unresolved till now. A new wave of interest to
such transform has been sparked by the recent development of the
Thermoacoustic Tomography (TAT or TCT) as one of the promising
methods of medical imaging (e.g.,
\cite{Kruger},\cite{MXW}-\cite{XWAK}). The TAT procedure can be
described as follows: a short microwave or radiofrequency
electromagnetic pulse is sent through the biological object. At
each internal location $x$ certain energy $H(x)$ is absorbed. The
cancerous cells happen to absorb several times more MW (or RF)
energy than the normal ones, which means that significant
increases of the values of $H(x)$ are expected at tumorous
locations. The absorbed energy, due to resulting heating, causes
thermoelastic expansion, which in turn creates a pressure wave.
This wave can be detected by ultrasound transducers placed outside
the object. It has been shown that here one effectively measures
the integrals of $H(x)$ over all spheres centered at the
transducers' locations. In other words, one needs to invert the
mentioned above generalized Radon transform of $H$
(``generalized,'' since integration is done over spheres). It is
clear from the dimension considerations that it should be
sufficient to run the transducers along a curve in the case of a
$2D$ problem or a surface in $3D$. The most popular geometries of
these surfaces (curves) that have been implemented are spheres,
planes, and cylinders \cite{MXW}-\cite{YXW2}.

The central problems that arise in these studies are:
uniqueness of reconstruction, reconstruction formulas and algorithms,
stability of the reconstruction, description of the range of the transform
and incomplete data problems.

All these questions have been essentially answered for the
classical Radon transform that arises in X-ray CT, Positron
Emission Tomography (PET), and Magnetic Resonance Imaging (MRI)
\cite{Natt4,Natt2001}. However, they are much more complex and not
that well understood for the circular Radon transform that arises
in TAT.

This paper contains some new approaches and results concerning the
uniqueness problem. The reader should be aware that for the
currently practically used geometries of TAT the uniqueness issue
has been resolved. E.g., for the spherical location of the centers
(transducers) uniqueness follows for instance from Corollary
\ref{C:boundary}, first proven in \cite{Kuch} (see also
\cite{ABK,AQ} and references therein). For the planar location, it
has been known for a long time \cite{John, CH} that only odd
functions with respect to this plane cannot be reconstructed from
the spherical integral data. However, the complete understanding
of the uniqueness problem for general locations of the transducers
remains elusive (especially in dimensions higher than two). The
aim of this paper is to make progress in filling this gap by
obtaining new uniqueness results, as well as by reproving some
known results by simpler means, which makes them easier to extend
to higher dimensions and other geometries.

The results of this paper were presented at the special sessions
on tomography at the AMS Meetings in Binghamton, NY in October
2003 and in Lawrenceville, NJ in April 2004 and at the Inverse
problems workshop at IPAM in November 2003.

The next section contains the mathematical formulation of the
problem and its brief history. The following section contains the main results of
this paper. It is followed by sections containing further remarks
and acknowledgements.

\section{Formulation of the problem and known results}

The discussion of the previous section motivates the study of the
following ``circular'' Radon transform. Let $f(x)$ be a continuous
function on $\Bbb{R}^n$, $n\ge 2$.

\begin{definition}\label{D:circular}The circular Radon transform of $f$ is defined
as
$$
Rf(p,r)=\int_{|y-p|=r}f(y)d\sigma(y),
$$
where $d\sigma(y)$ is the surface area on the sphere $|y-p|=r$
centered at $p \in \Bbb{R}^n$.
\end{definition}

In this definition we do not restrict the set of centers $p$ or
radii $r$. It is clear, however, that this mapping is
overdetermined, since the dimension of pairs $(p,r)$ is $n+1$,
while the function $f$ depends on $n$ variables only. This
suggests to restrict the set of centers to a set (hypersurface) $S
\subset \Bbb{R}^n$, while not imposing any restrictions on the
radii. We denote this restricted transform by $R_S$:
$$
R_Sf(p,r)=Rf(p,r)|_{p \in S}.
$$

\begin{definition}
The transform $R$ is said to be {\bf injective} on a set $S$ ($S$
is a {\bf set of injectivity}) if for any $f\in C_c(\Bbb{R}^n)$
the condition $Rf(p,r)=0$  for all $r\in \Bbb{R}$ and all $p\in S$
implies $f\equiv0$.

In other words, $S$ is a set of injectivity, if the mapping $R_S$
is injective on $C_c(\Bbb{R}^n)$.
\end{definition}
Here we use the standard notation $C_c(\Bbb{R}^n)$ for the space
of compactly supported continuous functions on $\Bbb{R}^n$.
The situation can be significantly different and harder to study
without compactness of support (or at least some decay) condition
\cite{ABK,AQ}. Fortunately, tomographic problems usually yield
compactly supported functions.

One now arrives to the

\begin{problem}\label{P:injec}
Describe all sets of injectivity for the circular Radon transform
$R$ on $C_c(\Bbb{R}^n)$.
\end{problem}

This problem has been around in different guises for quite a while
\cite{AQ,Leon,LP1,LP2}. The paper \cite{AQ} contains a survey of
some other problems that lead to the injectivity question for
$R_S$.

The first important observations concerning non-injectivity sets
were made by V.~Lin and A.~Pincus \cite{LP1,LP2} and by N.~Zobin
\cite{Zob}. Their results imply in particular that if $R$ is not
injective on $S$, then $S$ is contained in the zero set of a
harmonic polynomial. Therefore we get a sufficient condition for
injectivity:

\begin{corollary}\label{C:harmonic_zero}
Any set $S\subset \Bbb{R}^n$ of uniqueness for the harmonic
polynomials is a set of injectivity for the transform $R$.
\end{corollary}
In particular, this implies
\begin{corollary}\label{C:boundary} If $U \subset \Bbb{R}^n$ is
a bounded domain, then $S=\partial U$ is a injectivity set of
$R$.
\end{corollary}
We will see later a different proof of this fact that does not use
harmonicity.

So, what are possible non-injectivity sets? Any hyperplane $S$ is
such a set. Indeed, for any function $f$ that is odd with respect
to $S$, one gets $R_S f \equiv 0$. There are other options as
well. In order to describe them in $2D$, let us first introduce
the following definition.
\begin{definition}\label{D:Coxeter}
For any $N\in \Bbb{N}$ denote by $\Sigma_N$ the Coxeter system of
$N$ lines $L_0, \dots, L_{n-1}$ in the plane\footnote{In the
formula below we identify the plane with the complex plane $\CC$.
}:
$$
L_k=\{te^{i\pi k/n}| -\infty<t<\infty\}.
$$
\end{definition}
In other words, $\Sigma_N$ is a ``cross'' of $N$ lines passing
through the origin and forming equal angles $\pi/N$. It is rather
easy to construct a non-zero compactly supported function that is
simultaneously odd with respect to all lines of a given Coxeter
set. Hence, $\Sigma_N$ is a non-injectivity set as well. Applying
any rigid motion $\omega$, one preserves non-injectivity property.
It has been also discovered that one can add any finite set $F$
preserving non-injectivity. Thus, all sets $\omega \Sigma_N \cup
F$ are non-injectivity sets. It was conjectured by V.~Lin and
A.~Pincus that these are the only non-injectivity sets for
compactly supported functions on the plane. Proving this
conjecture, M.~Agranovsky and E.~Quinto established the following
result:
\begin{theorem}\label{T:AQ}\cite{AQ}
The following condition is necessary and sufficient for a set
$S\subset \Bbb{R}^2$ to be a set of injectivity for the circular
Radon transform on $C_c(\Bbb{R}^2)$:

$S$ is not contained in any set of the form
$\omega(\Sigma_N)\bigcup F$, where $\omega$ is a rigid motion in
the plane and $F$ is a finite set.
\end{theorem}

The (unproven) conjecture below describes non-injectivity sets in
higher dimensions.

\begin{conjecture}\label{C:conj}\cite{AQ}
The following condition is necessary and sufficient for $S$ to be
a set of injectivity for the circular Radon transform on
$C_c(\Bbb{R}^n)$:

$S$ is not contained in any set of the form $\omega(\Sigma)\bigcup
F$, where $\omega$ is a rigid motion of $\Bbb{R}^n$, $\Sigma$ is
the zero set of a {\bf homogeneous} harmonic polynomial, and $F$
is an algebraic subset in $\Bbb{R}^n$ of co-dimension at least
$2$.
\end{conjecture}
The reader notices that for $n=2$ this boils down to Theorem
\ref{T:AQ}.

The beautiful proof of Theorem \ref{T:AQ} by M. Agranovsky and E.
Quinto is built upon the following tools: microlocal analysis
(Fourier Integral Operators technique) that guarantees existence
of certain analytic wave front sets at the boundary of the support
of a function located on one side of a smooth surface (Theorem
8.5.6 in \cite{Hor}), and known geometric structure of level sets
of harmonic polynomials in $2D$ (e.g., \cite{Flatto}). These
methods, unfortunately, restrict wider applicability of the proof.
The microlocal tool works at an edge of the support and hence is
not applicable for non-compactly-supported functions. On the other
hand, the geometry of level sets of harmonic polynomials does not
work well in dimensions higher than 2 or on more general
Riemannian manifolds (e.g., on the hyperbolic plane). Thus, the
quest has been active for alternative approaches since \cite{AQ}
has appeared.

It is instructive to look at alternative reformulations of the problem
(which there are plenty \cite{AQ}). There is a revealing
reformulation \cite{AQ,Kuch} that stems from known relations
between spherical integrals and the wave equation (e.g., \cite{CH,
John}). Namely, consider the initial value problem for the wave
equation in $\Bbb{R}^n$:
\begin{equation}
u_{tt}-\triangle u=0, \;\; x\in \Bbb{R}^n, t\in \Bbb{R}
\label{E:wave1}
\end{equation}
\begin{equation}
u|_{t=0}=0, \;\; u_t|_{t=0}=f.\label{E:wave2}
\end{equation}
Then
$$
u(x,t)=\frac{1}{(n-2)!}\frac{\partial^{n-2}}{\partial
t^{n-2}}\int_0^t r(t^2-r^2)^{(n-3)/2}(Rf)(x,r)dr, \;\;t\ge0.
$$
Hence, it is not hard to show \cite{AQ} that the original problem
is equivalent to the problem of recovering $u_t(x,0)$ from the
value of $u(x,t)$ on subsets of $S\times(-\infty,\infty)$.
\begin{lemma}\label{L:wave}\cite{AQ,Kuch} A set $S$ is a
non-injectivity set for $C_c(\Bbb{R}^n)$ if and only if there
exists a non-zero compactly supported continuous function $f$ such
that the solution $u(x,t)$ of the problem
(\ref{E:wave1})-(\ref{E:wave2}) vanishes for any $x \in S$ and any
$t$.
\end{lemma}
Hence, non-injectivity sets are exactly the nodal sets of
oscillating free infinite membranes. In other words, injectivity
sets are those that observing the motion of the membrane over $S$
gives complete information about the motion of the whole membrane.

One can now try to understand the geometry of non-injectivity sets
in terms of wave propagation. The first example of such a consideration
was the original proof \cite{Kuch} of Corollary \ref{C:boundary} that
did not use harmonicity (not known at the time). Let $S=\partial U$ be
a non-injectivity (and hence nodal for wave equation) set, where $U$
is a bounded domain. Then on one hand, the membrane is free and hence
the energy of the initial compactly supported perturbation must move
away. Thus, its portion inside $U$ should decay to zero. On the other
hand, one can think  that $S$ is a fixed boundary and hence the
energy inside must stay constant. This contradiction allows one to
conclude that in fact $f=0$. The same PDE idea, with many more
technical details, was implemented in \cite{ABK} to prove the
following statement:
\begin{theorem}\label{T:ABK}\cite{ABK} If $U$
is a bounded domain in $\Bbb{R}^n$, then $S=\partial U$ is an
injectivity set for $R$ in the space $L^q(\Bbb{R}^n)$ if $q \leq
2n/(n-1)$. This statement fails when $q > 2n/(n-1)$, in which case
spheres fail to be injectivity sets.
\end{theorem}

In spite of these limited results, it still had remained unclear
what distinguishes in terms of wave propagation the ``bad'' flat
lines $S$ in Theorem \ref{T:AQ} that can be nodal for all times,
from any truly curved $S$ that according to this theorem cannot
stay nodal. An approach to this question was found in the recent
paper \cite{FPR} by D.~Finch, Rakesh, and S.~Patch, where in
particular some parts of the injectivity results due to \cite{AQ}
were re-proven by simple PDE means without using microlocal tools
and harmonicity:

\begin{theorem}\label{T:Finch}\cite{FPR}
Let $D$ be a bounded, open, subset of $\Bbb{R}^n$, $n\ge2$, with a
strictly convex smooth boundary $S$. Let $\Gamma$ be any
relatively open subset of $S$. If $f$ is a smooth function on
$\Bbb{R}^n$ supported in $\bar{D}$, $u$ is the solution of the
initial value problem (1), (2) and $u(p,t)=0$ for all $t$ and
$p\in\Gamma$, then $f=0$.
\end{theorem}

Although this theorem follows from microlocal results in
\cite{AQ}\footnote{Results of \cite{AQ} make the situation
described in Theorem \ref{T:Finch} impossible, since the support
of $f$ lies on one side of a tangent plane to $\Gamma$. See also
Theorem \ref{tangent} and \cite{LQ}.}, its significance lies in
the proof provided in \cite{FPR} (that paper contains important
results concerning inversion as well, which we do not touch here).

The following two standard statements concerning the unique
continuation and finite speed of propagation for the wave equation
were the basis of the proof of the Theorem \ref{T:Finch} in
\cite{FPR}. They will be relevant for our purpose as well.

\begin{proposition}\label{P:uniqueness_cont}\cite{FPR}
Let $B_\epsilon(p)=\{x\in \Bbb{R}^n \, |\, |x-p|<\epsilon\}$. If
$u$ is a distribution and satisfies (\ref{E:wave1}) and $u$ is
zero on $B_\epsilon(p)\times(-T,T)$ for some $\epsilon>0$, and
$p\in \Bbb{R}^n$, then u is zero on
$$
\{(x,t):|x-p|+|t|<T\},
$$
and in particular on
$$
\{(x,0):|x-p|<T\}.
$$
\end{proposition}

Let now $D$ be a bounded, open subset of $\Bbb{R}^n$ with the
boundary $S$. For points $p,q$ outside $D$, let $d(p,q)$ denote
the infimum of the lengths of all the piecewise $C^1$ paths in
$\Bbb{R}^n\setminus D$ joining $p$ to $q$. Then $d(p,q)$ is a
metric on $\Bbb{R}^n\setminus D$. For any point $p$ in
$\Bbb{R}^n\setminus D $ and any positive number $r$, define
$E_r(p)$ to be the ball of radius $r$ and center at $p$ in
$\Bbb{R}^n\setminus D$ with respect to this metric, i.e.
$$
E_r(p)=\{x\in \Bbb{R}^n\setminus D:d(x,p)<r\}.
$$

\begin{proposition}\label{P:domain_dep}\cite{FPR}
Suppose $D$ is a bounded, open, connected subset of $\Bbb{R}^n$,
with a smooth boundary $S$. Let $u$ be a smooth solution of the
exterior problem
$$
u_{tt}-\triangle u=0, \;\;x\in \Bbb{R}^n\setminus D,\;\;t\in R
$$
$$
u=h\;\; on \;\; S\times R.
$$
Suppose $p$ is not in $D$, and $t_0<t_1$ are real numbers. If
$u(.,t_0)$ and $u_t(.,t_0)$ are zero on $E_{t_1-t_0}(p)$ and $h$
is zero on
$$
\{(x,t):x\in S, t_0\le t\le t_1, d(x,p)\le t_1-t \},
$$
then $u(p,t)$ and $u_t(p,t)$ are zero for all $t\in [t_0,t_1)$.
\end{proposition}


\section{Further injectivity results by PDE means}\label{S:main}

We will show now how simple tools similar to the Propositions
\ref{P:uniqueness_cont} and \ref{P:domain_dep}, namely finite
speed of propagation and domain of dependence for the wave
equation allow one to obtain more results concerning geometry of
non-injectivity sets, as well as to re-prove some known results
with much simpler means. The final goals were to recover the full
result of \cite{AQ} in $2D$ and to prove its analogs in higher
dimensions and for other geometries (e.g., hyperbolic one) using
these simple means. Albeit this goal has not been completely
achieved yet, we can report some progress in all these directions.

Let us start with some initial remarks that will narrow the cases
we need to consider. First of all, one can assume functions $f$ as
smooth as we wish, since convolution with smooth radial mollifiers
does not change the fact that $R_S f=0$ (e.g., \cite{AQ}).
Secondly, according to the results mentioned before, any
non-injectivity set $S$ in the class of compactly supported
functions is contained in an algebraic surface that is also a
non-injectivity set. It is rather straightforward to show that the
same is true for functions that decay exponentially. Thus, {\bf
considering only exponentially decaying functions, one does not
restrict generality by assuming from the start algebraicity of
$S$}. It is known \cite{A} that algebraic surfaces of co-dimension
higher than 1 are automatically non-injectivity sets. Thus, we can
restrict our attention to algebraic hypersurfaces $S$ of
$\Bbb{R}^n$ only. Any set that is not algebraic (or rather, is not
a part of such an algebraic surface) is automatically an
injectivity set. So, when trying to obtain necessary conditions
for non-injectivity, confining ourselves to the case of algebraic
hypersurfaces solely we do not lose any generality. One can also
assume irreducibility of that surface, if this helps. When needed,
one can also exclude the case of closed hypersurfaces, since
according to Corollary \ref{C:boundary} those are all injectivity
sets.

Our goal now is to exclude some pairs $(S,f)$, where $S$ is an
algebraic surface and $f$ is a non-zero function as possible
candidates for satisfying the non-injectivity condition $R_Sf=0$.
We will do this in terms of geometry of the support of function
$f$. Notice that Theorem \ref{T:Finch} does exactly that when $S$
contains an open part of the boundary of a smooth strictly convex
domain where $f$ is supported. Theorem \ref{T:AQ}, on the other
hand excludes all compactly supported $f$'s in $\Bbb{R}^2$, unless
$S=\omega\Sigma_N$. Similarly, Theorem \ref{T:ABK} excludes
boundaries $S$ of bounded domains when $f$ is in an appropriate
space $L_p(\Bbb{R}^n)$.

Let $S$ be an algebraic hypersurface (which can be assumed to be
irreducible if needed) that splits $\Bbb{R}^n$ into connected
parts $H^j$, $j=1, ..., m$. One can define the interior metric in
$H^j$ as follows:
\begin{equation}\label{E:metric}
d^j(p,q)=inf\{\mbox{length of}\, \gamma\},
\end{equation}
where the infimum is taken over all $C^1$-curves $\gamma$ in $H^j$
joining points $p,q \in H^j$.

\begin{theorem}\label{T:halves}
Let $S$ and $H^j$ be as above and $f\in C(\Bbb{R}^n)$ be such that
$R_Sf=0$. Let also $x\in \bar{H^j}$, where $\bar{H^j}$ is the
closure of $H^j$. Then
\begin{equation}\label{E:eqhalves_ineq}
\begin{array}{c}
  dist(x, \mbox{supp } f \cap H^j)=dist^j(x, \mbox{supp } f \cap H^j)\\
  \leq dist(x, \mbox{supp } f \cap H^k),\, k \neq j,
\end{array}
\end{equation}
where distances $dist^j$ are computed with respect to the metrics
$d^j$, while $dist$ is computed with respect to the Euclidean
metric in $\Bbb{R}^n$.

In particular, for  $x\in S$ and any $j$
\begin{equation}\label{E:eqhalves_eq}
\begin{array}{c}
  dist(x, \mbox{supp } f \cap H^j)=dist^j(x, \mbox{supp } f \cap H^j)
  = dist(x, \mbox{supp } f ).
\end{array}
\end{equation}
Thus, the expressions in (\ref{E:eqhalves_eq}) in fact do not
depend on $j=1, ..., m$.
\end{theorem}
\begin{remark}
Notice that under the condition of algebraicity of $S$ the theorem
does not require the function $f$ to be compactly supported and in
fact imposes no condition on behavior of $f$ at infinity. On the
other hand, as it has been mentioned before, if $f$ decays
exponentially, then the algebraicity assumption does not restrict
the generality of consideration.
\end{remark}
\noindent {\bf Proof of the theorem.} Notice first of all, that
the function $d^j(p,x)$ has gradient $|\nabla_x d^j(p,x)|\leq 1$
a.e.\footnote{In order to justify legality of the calculation
presented below, one can either use geometric measure theory
tools, as in \cite{FPR}, or just notice that due to algebraicity
of $S$, the function $d^j(p,x)$ is piece-wise analytic.}

Let us prove now the equality
\begin{equation}\label{E:onehalf}
   dist(x, \mbox{supp } f \cap H^j)=dist^j(x, \mbox{supp } f \cap
   H^j).
\end{equation}
Since $d^j(p,q)\geq |p-q|$, it is sufficient to
prove that the left hand side expression cannot be strictly
smaller than the one on the right. Assume the opposite, that
\begin{equation}\label{E:hal_opposite1}
dist(x, \mbox{supp } f \cap H^j)=d_1<d_2=dist^j(x, \mbox{supp } f
\cap
   H^j).
\end{equation}
Pick a smaller segment $[d_3,d_4]\subset (d_1,d_2)$. Then, by
continuity, for any point $p$ in a small ball $B \subset H^j$ near
$x$ (not necessarily containing $x$, for instance when $x \in S$)
one has
\begin{equation}\label{E:hal_opposite}
dist(p, \mbox{supp } f \cap H^j)\leq d_3<d_4\leq dist^j(p,
\mbox{supp } f \cap
   H^j).
\end{equation}
For such a point $p$, consider the volume $V$ in the space-time
region $H^j\times\Bbb{R}$ bounded by the space-like surfaces
$\Sigma_1$ given by $t=0$ and $\Sigma_2$ described as $t=\phi
(x)=\tau-d^j(p,x)$, and the ``vertical'' boundary $S\times
\Bbb{R}$. Here $\tau \leq (d_3+d_4)/2$. Consider the solution
$u(x,t)$ of the wave equation problem
(\ref{E:wave1})--(\ref{E:wave2}) with the initial velocity $f$.
Then, by construction, this solution and its time derivative are
equal to zero at the lower boundary $t=0$ and on the lateral
boundary $S\times\Bbb{R}$. Hence, by the standard energy
computation (integrating the equality $u \Box u =0$, see, e.g.,
Section 2.7, Ch. 1 in \cite{BJS}) we conclude that $u=0$ in $V$.
For the reader's convenience, let us provide brief details of the
corresponding calculations from \cite{BJS}: Since $\Box u=0$,
$u=u_t=0$ on $\Sigma_1$, and $u|_S=0$ for all times, we get by
integration by parts
\begin{equation}\label{E:BJS}
\begin{array}{c}
    0=\int \limits_V u_t \Box u dxdt=\int\limits_{t=\phi(x)} \frac12
    \left(|\nabla u|^2+u_t^2+2u_t\nabla \phi \cdot \nabla u
    \right)dx\\
    = \frac12\int\limits_{\phi(x)\geq 0}
    \left(|\nabla (u(x,\phi (x))|^2+(1-|\nabla \phi|^2)u_t(x,\phi (x))^2
    \right)dx.
\end{array}
\end{equation}
Since $|\nabla \phi| \leq 1$, we conclude that
$$
\int\limits_{\phi(x)\geq 0} \left(|\nabla (u(x,\phi (x))|^2
\right)dx =0
$$
and hence $u$ is constant on $\Sigma_2$. Taking into the account
the zero conditions on $S$ and $\Sigma_1$, one concludes that
$u=0$ on $\Sigma_2$, and hence in $V$.

In particular, $u(p,t)=0$ for all $p\in B$ and $|t|\leq
(d_3+d_4)/2$. Notice that $(d_3+d_4)/2>d_3$. Now applying
Proposition \ref{P:uniqueness_cont} to the wave equation in the
whole space, we conclude that
\begin{equation}\label{E:wholespace}
dist(p,\mbox{supp }f )>d_3,
\end{equation}
and hence
\begin{equation}\label{E:almostthere}
  dist(p,\mbox{supp }f \cap H^j)>d_3,
\end{equation}
which is a contradiction. This proves (\ref{E:onehalf}). It is now
sufficient to prove
\begin{equation}\label{E:equals}
  dist(x,\mbox{supp }f \cap H^j)\leq dist(x,\mbox{supp }f \cap H^k)
\end{equation}
for $k \neq j$. This in fact is an immediate consequence of
(\ref{E:wholespace}). Alternatively, we can repeat the same
consideration as above in a simplified version. Namely, suppose
that
\begin{equation}\label{E:nonequalhalves}
  dist(x,\mbox{supp }f \cap H^j)> d_2>d_1> dist(x,\mbox{supp }f \cap H^k)
\end{equation}
for a point $x \in H^j \cap S$, and hence for all points $p$ in a
small ball in $H^j$. Consider the volume $V$ in the space-time
region $H^j\times\Bbb{R}$ bounded by the space-like surfaces $t=0$
and $t=d_2-|x-p|$ ($p$ fixed in the small ball) and the boundary
$S\times \Bbb{R}$. Consider the solution $u(x,t)$ of the wave
equation problem (\ref{E:wave1})-(\ref{E:wave2}) with the initial
velocity $f$. Then, by construction, this solution and its time
derivative are equal to zero at the lower boundary $t=0$ and on
the lateral boundary $S\times\Bbb{R}$. Hence, by the same standard
domain of dependence argument (see, e.g., Section 2.7, Ch. 1 in
\cite{BJS}) we conclude that $u=0$ in $V$. In particular,
$u(p,t)=0$ for all $p\in B$ and $|t|\leq d_2$. Now applying
Proposition \ref{P:uniqueness_cont} to the wave equation in the
whole space, we conclude that
$$
dist(p,\mbox{supp }f )>d_2,
$$
and hence
\begin{equation}
  dist(p,\mbox{supp }f \cap H^k)>d_2,
\end{equation}
which is a contradiction. \noindent $\Box$

We will now show several corollaries that can be extracted from
Theorem \ref{T:halves}.

\begin{corollary}\label{C:perpendicular_ray}
Let $f$ be continuous and $S\subset \Bbb{R}^n$ be an algebraic
hypersurface such that $R_Sf=0$. Let $L$ be any hyperplane such
that $L\cap \mbox{supp }f\neq \emptyset$ and such that $\mbox{supp
}f$ lies on one side of $L$. Let $x\in L\cap \mbox{supp }f$ and
$r_x$ be the open ray starting at $x$, perpendicular to $L$, and
going into the direction opposite to the support of $f$. Then
either $r_x \subset S$ (and hence, the whole line containing $r_x$
belongs to $S$), or $r_x$ does not intersect $S$.
\end{corollary}
{\bf Proof.} Assuming otherwise, let $p\in r_x \cap S$ and $H^j$
be the connected components of $\RR^n \backslash S$ such that $p$
belongs to their closures. Since $x$ is the only closest point to
$p$ in the support of $f$, Theorem \ref{T:halves} implies that for
any $j$ there exist paths $t_\epsilon$ joining $x$ and $p$ through
$H^j$ and such that the length of $t_\epsilon$ tends to $|x-p|$
when $\epsilon \to 0$. This means that these paths converge to the
linear segment $[x,p]$. Hence, this segment belongs to $H^j$ for
any $j$, and thus to $\mathop{\cap}\limits_j H^j$, which is a part
of $S$. We conclude that the segment $[x,p]$, and then, due to
algebraicity of $S$, the whole its line belongs to $S$. This
proves the statement of the corollary. $\Box$

One notices that a similar proof establishes the following
\begin{corollary}\label{C:closest}
Let $f$ be continuous and $S\subset \Bbb{R}^n$ be an algebraic
hypersurface such that $R_Sf=0$. Suppose $p\in S$ is such that $p$
does not belong to $\mbox{supp }f$ and there exists unique point
$x$ in $\mbox{supp }f$ closest to $p$. Then $S$ contains the whole
line passing through the points $x$ and $p$.
\end{corollary}

Let  $S\subset \Bbb{R}^n$. For any points $p,q \in \Bbb{R}^n-S$ we
define the distance $d_S(p,q)$ as the infimum of lengths of $C^1$
paths in $\Bbb{R}^n-S$ connecting these points. Clearly
$d_S(p,q)\geq |p-q|$. Using this metric, we can define the
corresponding distances $dist_S$ from points to sets.

\begin{theorem}\label{T:piece}
Let a set $S\subset \Bbb{R}^n$ and a non-zero function $f \in
C(\Bbb{R}^n)$ exponentially decaying at infinity be such that
$R_Sf=0$. Then for any point $p\in \Bbb{R}^n-S$
\begin{equation}\label{E:piece}
  dist_S(p,\mbox{supp} f)=dist(p,\mbox{supp} f).
\end{equation}
The same conclusion holds for any continuous function, if one
assumes that $S$ is an algebraic hypersurface.
\end{theorem}

{\bf Proof.} Assume that (\ref{E:piece}) does not hold, i.e.
$$
dist_S(p,\mbox{supp} f)>dist(p,\mbox{supp} f).
$$
As it has been mentioned before, under the conditions of the
theorem, we can assume $S$ to be a part of an algebraic surface
$\Sigma$ for which $R_\Sigma f=0$. Let $\Sigma$ divide the space
into parts $H^j$. Then, in notations of the previous theorem, we
have
\begin{equation}
dist^j(p,\mbox{supp } f\cap H^j) \geq dist_S(p,\mbox{supp} f)
\end{equation}
and hence
\begin{equation}
  dist^j(p,\mbox{supp } f\cap H^j) >dist(p,\mbox{supp} f).
\end{equation}
 This, however, contradicts Theorem \ref{T:halves}. \noindent $\Box$

Let us formulate another example of a geometric constraint on
pairs $S,\,f$ such that $R_Sf=0$.\footnote{A similar statement in
the case of analytic surfaces $S$ was announced in \cite{LQ} for
distributions $f$. The proof is claimed to be based upon
microlocal analysis.}
\begin{theorem}\label{tangent}
Let $S\subset \Bbb{R}^n$ be a relatively open piece of a
$C^1$-hypersurface and $f\in C_c(\RR^n)$ be such that $R_Sf=0$. If
there is a point $p_0\in S$ such that the support of $f$ lies
strictly on one side of the tangent plane $T_{p_0}S$ to $S$ at
$p_0$, then $f=0$.\footnote{This implies, in particular, Theorem
\ref{T:Finch}.}
\end{theorem}
{\bf Proof of the theorem.} Let us denote by $K_p(\mbox{supp }f)$
the convex cone with the vertex $p$ consisting of all the rays
starting at $p$ and passing through the convex hull of the support
of $f$. Then $K_{p_0}(\mbox{supp }f)$, due to the condition of the
theorem, lies on one side of $T_{p_0}S$, touching it only at the
point $p_0$. Let us pull the point $p_0$ to the other side of the
tangent plane along the normal to a nearby position $p$. Then it
is easy to see that for $p$ sufficiently close to $p_0$, all rays
of the cone $K_{p}(\mbox{supp }f)$ will intersect $S$. This means
in particular, that for this point $p$ we have $dist_S(p,
\mbox{supp }f)>dist(p, \mbox{supp }f)$. According to Theorem
\ref{T:piece}, this implies that $f=0$. $\Box$

\begin{corollary}\label{C:intersect}
Let $S\subset \Bbb{R}^n$ be an algebraic hypersurface and $f\in
C_c(\RR^n)$. If $R_Sf=0$, then every tangent plane to $S$
intersects the convex hull of the support of $f$.
\end{corollary}

The above results present significant restrictions on the geometry
of the non-injectivity sets $S$ and of the supports of functions
$f$ in the kernel of $R_S$. One can draw more specific conclusions
about these sets. The statement below was proven in \cite{AQ} by
using the geometry of zeros of harmonic polynomials, which we
avoid.

\begin{proposition}\label{P:asymptotes}
Let $S\subset \Bbb{R}^2$ be an algebraic curve such that $R_Sf=0$
for some non-zero compactly supported continuous function $f$.
Then $S$ has no compact components, and each its component has
asymptotes at infinity.
\end{proposition}
{\bf Proof.} Corollary \ref{C:boundary} excludes bounded
components. So, we can think that $S$ is an irreducible unbounded
algebraic curve. Existence of its asymptotes can be shown as
follows. Let us take a point $p\in S$ and send it to one of the
infinite ends of $S$. According to Corollary \ref{C:intersect},
every tangent line $T_pS$ intersects the convex hull of the
support of $f$, which is a fixed compact in $\Bbb{R}^2$. This
makes this set of lines on the plane compact. Hence, we can choose
a sequence of points $p_j$ such that the lines $T_{p_j}S$ converge
to a line $T$ in the natural topology of the space of lines (e.g.,
one can use normal coordinates of lines to introduce such
topology). This line $T$ is in fact the required asymptote.
Indeed, let us choose the coordinate system where $T$ is the
$x$-axis. Then the slopes of the sequence $T_{p_j}S$ converge to
zero. Due to algebraicity, for a tail of this sequence, the
convergence is monotonic, and in particular holds for all $p\in S$
far in the tail of $S$. Let us for instance assume that these
slopes are negative. Then the tail of $S$ is the graph of a
monotonically decreasing positive function. This means that $S$
has a horizontal asymptote. This asymptote must be the $x$-axis
$T$, otherwise the $y$-intercepts of $T_{p_j}S$ would not converge
to zero, which would contradict the convergence of $T_{p_j}S$ to
$T$. $\Box$

The next statement proves the Agranovsky-Quinto Theorem \ref{T:AQ}
in some particular cases. In order to formulate it, we need to
introduce the following condition:

{\bf Condition A.} Let $K$ be a compact subset of $\Bbb{R}^n$. We
will say that the boundary of $K$ satisfies {\bf condition
A}\footnote{This condition essentially restricts the curvature of
the boundary from below. }, if there exists a positive number
$r_0$ such that for any $r<r_0$ and any point $x$ in the infinite
connected component of $\Bbb{R}^n \setminus K$ such that
$\mbox{dist }(x,K)=r$ there exists a unique point $k$ on $K$ such
that $|x-k|=r$.

Examples of such sets are convex sets (where $r_0>0$ can be chosen
arbitrarily) and sets with a $C^2$ boundary (where $r_0$ should be
sufficiently small).
\begin{theorem}\label{P:convex2D}
Let $S\subset \Bbb{R}^2$ and $f(\neq 0)\in C_c(\Bbb{R}^2)$ be such
that $R_Sf=0$. If the external boundary of the support of $f$
(i.e., the boundary of the infinite component of the complement of
the support) is connected and satisfies Condition A, then
$S\subset \omega \Sigma _N \cup F$ in notations of Theorem
\ref{T:AQ}.

The conditions of the theorem are satisfied for instance when the
support of $f$ contains the boundary of its convex hull, or when
the support's external boundary is connected and of the class
$C^2$.
\end{theorem}
{\bf Proof.} First of all, up to a finite set, we can assume that
$S$ is an algebraic curve. Since the external boundary of the
support is assumed to be connected, Theorem \ref{T:halves} implies
that any irreducible component of $S$ must meet any neighborhood
of the support of $f$. If we take the neighborhood of radius
$r<r_0$, then each point on $S$ in this neighborhood will have a
unique closest point on $\mbox{supp }f$. Applying now Corollary
\ref{C:closest}, we conclude that $S$ consists of straight lines
$L_j$ intersecting the support. It is known that any straight line
$L$ is a non-injectivity set, but the only functions annihilated
by $R_L$ are the ones odd with respect to $L$ (e.g.,
\cite{AQ,CH,John}). Hence, $f$ is odd with respect to all lines
$L_j$. In particular, every of these lines passes through the
center of mass of the support of $f$. Hence, lines $L_j$ form a
``cross''\footnote{One can prove that all these lines pass through
a joint point also in a different manner. Indeed, due to oddness
of $f$, each line is a symmetry axis for the support of $f$. Then,
considering the group generated by reflections through these
lines, one can easily conclude that if they did not pass through a
joint point, then the support of $f$ must have been non-compact.}.
It remains now to show that the angles between the lines are
commensurate with $\pi$. This can also be shown in several
different ways. For instance, this follows immediately from
existence of a {\bf harmonic} polynomial vanishing on $S$. Another
simple option is to notice that if this is not the case, then
there is no non-zero function that is odd simultaneously with
respect to all the lines. $\Box$

Exactly the same consideration as above shows that in higher
dimensions the following statement holds:
\begin{proposition}\label{P:convex3D}
Let $S\subset \Bbb{R}^n$ and $f(\neq 0)\in C_c(\Bbb{R}^n)$ be such
that $R_Sf=0$. If the external boundary of the support of $f$
(i.e., the boundary of the infinite component of the complement of
the support) is connected and satisfies Condition A, then $S$ is
ruled (a scroll)\footnote{A {\bf ruled surface}, or a {\bf scroll}
is the union of a family of lines (e.g., \cite{Spivak})}.

The conditions of the theorem are satisfied for instance when the
support of $f$ contains the boundary of its convex hull, or when
the support's external boundary is connected and of the class
$C^2$.
\end{proposition}
\begin{remark}
If we could also show that all these lines pass through the same
point, then this would immediately imply, as in the previous
proof, the validity of Conjecture \ref{C:conj} for this particular
case.
\end{remark}
\section{Additional remarks}
\begin{enumerate}
\item M.~Agranovsky and E.~T.~Quinto have written besides \cite{AQ},
several other
papers devoted to the problem considered here. They consider some
partial cases (e.g., distributions $f$ supported on a finite set)
and variations of the problem (e.g., in bounded domains rather
than the whole space). See \cite{A,AQ2,AQ3,AVZ} for details.

\item One of our goals was to obtain the complete Theorem
\ref{T:AQ}, the main result of \cite{AQ} by simple PDE tools,
avoiding using the geometry of zeros of harmonic polynomials and
microlocal analysis (or at least one of those), as well as to
prove its analogs in higher dimensions and for other geometries
(e.g., hyperbolic one). Although we have not completely succeeded
in this yet, the results presented (e.g., Propositions
\ref{P:asymptotes} and \ref{P:convex3D} and Theorem
\ref{P:convex2D}) are moving in this direction.

\item The PDE methods presented here in principle bear a potential
for considering non-compactly-supported functions. In order to
achieve this, one needs to have qualitative versions of statements
like Proposition \ref{P:domain_dep} and Theorem \ref{T:piece},
where instead of just noticing whether a wave has come to certain
point at a certain moment (which was our only tool), one controls
the amount of energy it carries.

\item In this paper one of the motivations for studying the injectivity
problem was the thermoacoustic tomography. One wonders then
if considerations of $2D$ problems (rather than $3D$ ones)
bear any relevance for TAT. In fact, they do. If either the
scanned sample is very thin, or the transducers are collimated in
such a way that they register the signals only coming parallel to
a given plane, one arrives to a $2D$ problem.

\item Most of our results can be generalized to some Riemannian
manifolds, in particular to the hyperbolic plane (where the analog
of Theorem \ref{T:AQ} has not been proven yet). We plan to address
these issues elsewhere. E.~T.~Quinto has recently announced a
version of Theorem \ref{tangent} in the case of distributions for
the spherical transforms on real-analytic Riemannian manifolds
with infinite injectivity radius and an analytic set $S$ of
centers \cite{Q2004}.

\item A closer inspection of the results of Section \ref{S:main} shows
that most of them have their local versions, where it is not
required that the whole transform $R_S$ of a function vanishes,
but rather only for radii up to a certain value. One can see an
example of a local uniqueness theorem for the circular transform
in \cite{LQ}. We hope to address this issue elsewhere.

\item As J.~Boman notified us during the April 2004 AMS meeting in
Lawrenceville, he jointly with J.~Sjostrand, being unaware of our
work, had recently independently obtained some results analogous
to some of those presented here (e.g., to Theorem \ref{T:piece}).

\item We have not touched the problem of finding explicit
inversion formulas for the circular transforms. Such formulas are
known for the spherical, planar, and cylindrical sets of centers
\cite{And, Den, Faw, FPR, Nil, Pal, MXW, YXW1, YXW2}. They come in
two kinds: the ones involving expansions into special functions,
and the ones of backprojection type. Exact backprojection type
formulas are known for the planar geometry \cite{Den,Pal} and
recently for the spherical geometry in odd dimensions \cite{FPR}
if the function to be reconstructed is supported inside the sphere
of transducers.

Another problem deserving attention is finding the ranges of
transforms $R_S$. Such knowledge could be used, for instance, to
replenish missing data. Some necessary range conditions have been
recently obtained in \cite{Patch} for spherical location of
transducers.

An important problem of reconstruction with
incomplete data was treated in \cite{LQ,XWAK} based on an earlier
work by E.~T.~Quinto in \cite{Q1993b}.


\item An important integral geometric technique of the so called
$\kappa$-operator has been developed in I.~Gelfand's school (e.g.,
\cite{GGG1,GGG2}). It has been applied recently to the problems of
the circular Radon transform (see \cite{Gi}, the last chapter of
\cite{GGG2}, and references therein), albeit applicability of this
method to the problems of the kind we consider in this paper is
not completely clear yet.
\end{enumerate}
\section{Acknowledgements}
The authors express their gratitude to M.~Agranovsky, J.~Boman,
E.~Chappa, M.~de~Hoop, L.~Ehrenpreis, D.~Finch, S.~Patch,
E.~T.~Quinto, L.~Wang, M.~Xu, Y.~Xu, and N.~Zobin for information
about their work and discussions. The authors are also grateful to
the reviewers for useful comments.

This research was partly based upon work supported by the NSF
under Grants DMS 0296150, 9971674, 0002195, and 0072248. The
authors thank the NSF for this support. Any opinions, findings,
and conclusions or recommendations expressed in this paper are
those of the authors and do not necessarily reflect the views of
the National Science Foundation.

\end{document}